\documentclass[10pt,aps,prb,twocolumn,superscriptaddress]{revtex4-1}
\usepackage{graphicx}
\usepackage{amssymb}
\usepackage{amsmath}
\usepackage{appendix}
\usepackage{comment}

\def\bj{\mathbf{j}}

\def\bn{\mathbf{n}}

\newcommand{\rf}[1]{(\ref{#1})}
\newcommand{\al}[1]{\begin{aligned}#1\end{aligned}}

\newcommand{\eq}[1]{\begin{equation}#1\end{equation}}

\begin{document}

\title{Collective spin dynamics under dissipative spin Hall torque}

\author{Yaroslav Tserkovnyak}
\affiliation{Department of Physics and Astronomy, University of California, Los Angeles, California 90095, USA}
\author{Eran Maniv}
\author{James Analytis}
\affiliation{Department of Physics, University of California, Berkeley, CA 94720, USA}
\affiliation{Materials Sciences Division, Lawrence Berkeley National Laboratory, Berkeley, California, 94720, USA}

\begin{abstract}
Current-induced spin torques in layered magnetic heterostructures have many commonalities across broad classes of magnetic materials. These include not only collinear ferromagnets, ferrimagnets, and antiferromagnets, but also more complex noncollinear spin systems. We develop a general Lagrangian-Rayleigh approach for studying the role of dissipative torques, which can pump energy into long-wavelength magnetic dynamics, causing dynamic instabilities. While the Rayleigh structure of such torques is similar for different magnetic materials, their consequences depend sensitively on the nature of the order and, in particular, on whether there is a net magnetic moment. The latter endows the system with a unipolar switching capability, while magnetically compensated materials tend to evolve towards limit cycles, at large torques, with chirality dependent on the torque sign. Apart from the ferromagnetic and antiferromagnetic cases, we discuss ferrimagnets, which display an intricate competition between switching and limit cycles. As a simple case for compensated noncollinear order, we consider isotropic spin glasses, as well as a scenario of their coexistence with a collinear magnetic order.
\end{abstract}

\maketitle

\textit{Introduction}|The spin Hall effect and the associated torque on magnetization dynamics encompass a vast array of nonequilibrium phenomena in diverse magnetic heterostructures with different microscopic origins.\cite{hoffmannIEEEM13,*sinovaRMP15} Here, we attempt to establish a common thread for thinking about the consequences of these torques on magnetic dynamics and switching in different families of magnetic materials: ferro-, ferri-, antiferro-magnets, and spin glasses. We focus on the most generic dissipative torques unconditioned by the crystalline symmetries. Through spin-orbit interactions, these are exerted by electrical currents on collective dynamics of magnetic degrees of freedom, steadily pumping energy into the latter.

While the formal structure of the theory is distilled down to universal ingredients, irrespective of the details of the magnetic order, the ensuing magnetic dynamics differ significantly for different families of the magnetic materials. In particular, we emphasize a general tendency for (anti)ferromagnetic correlations to align along (perpendicular) the effective spin-accumulation direction produced by the dissipative spin torque. Ferrimagnets exhibits a competition between these opposite propensities, while magnetically-compensated materials, such as spin glasses, share much commonality with antiferromagnets. The situation can become especially intriguing when there is a coexistence of competing magnetic orders in the same material.\\

\textit{Collinear order parameter.}|As a central generic model, we consider low-temperature classical dynamics of an ordered collinear \textit{ferri}magnet,\cite{ivanovJETP83} in the presence of Gilbert damping\cite{gilbertIEEEM04} and spin Hall torque.\cite{tserkovPRB14} Following Ref.~\onlinecite{tserkovPRB17}, we write the magnetic Lagrangian density $\mathcal{L}[\mathbf{n},\mathbf{m}]$ (focusing on the dominant kinetic and energy terms) as
\eq{
\mathcal{L}=-s\,\mathbf{a}(\mathbf{n})\cdot\partial_t\mathbf{n}+\mathbf{m}\cdot\mathbf{n}\times\partial_t\mathbf{n}-\frac{\mathbf{m}^2}{2\chi}+(\mathbf{m}+s\mathbf{n})\cdot\mathbf{b}-\mathcal{E}(\mathbf{n})\,.
\label{L}}
The collective dynamics are parametrized by the directional order parameter $\mathbf{n}(t)$ (s.t., $|\mathbf{n}|\equiv1$) and a small transverse spin density $\mathbf{m}(t)$ (obeying the constraint $\mathbf{m}\cdot\mathbf{n}\equiv0$ and realizing generators for the order-parameter rotations\cite{andreevSPU80}). $s$ is the (uncompensated) equilibrium longitudinal spin density along the order parameter (zero in the purely antiferromagnetic limit) and $\chi\propto J^{-1}$ is the transverse spin susceptibility (where $J$ is the microscopic Heisenberg exchange: assumed to be the largest energy scale in the problem). The first term is the ferro-like (Wess-Zumino) kinetic term, expressed in terms of a vector potential $\mathbf{a}(\mathbf{\mathbf{n}})$ produced on a unit sphere by a magnetic monopole of unit charge. The second term is the antiferro-like $\sigma$-model kinetic term. Both of these kinetic terms stem from the Berry phases summed over the individual spins.\cite{auerbachBOOK94} The remaining terms consist of Zeeman energy proportional to $\mathbf{b}\equiv\gamma\mathbf{B}$, in terms of the gyromagnetic ratio $\gamma$ and magnetic field $\mathbf{B}$, and energy $\mathcal{E}(\mathbf{n})$ that includes all other order-parameter-dependent terms, such as dipolar interactions (when $s\neq0$), anisotropies, and exchange-stiffness terms (in the case of order-parameter inhomogeneities).

\begin{figure}[!t]
\begin{center}\includegraphics[width=\linewidth]{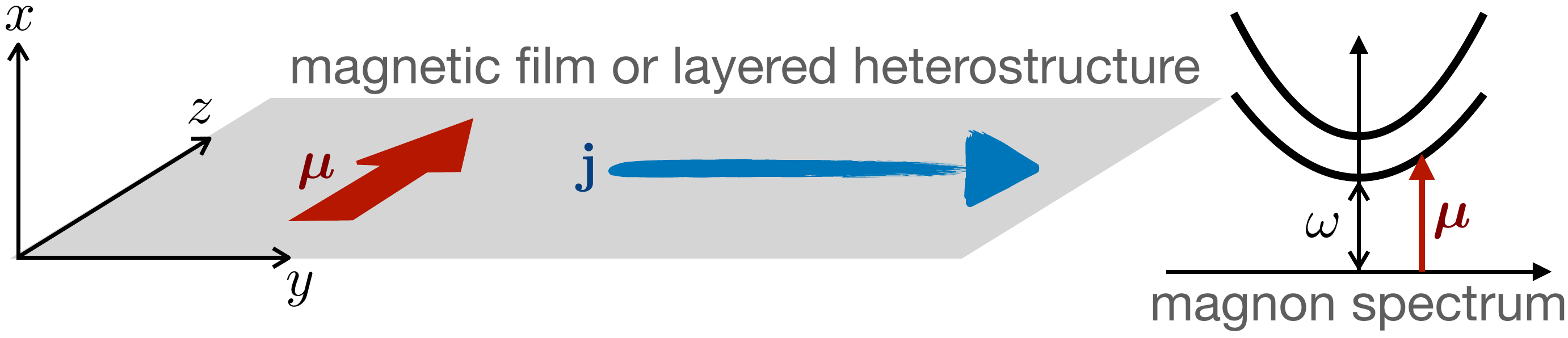}\end{center}
\caption{The schematic of a current-biased planar structure: The magnetic film is in the $yz$ plane. The reflection symmetry is broken along the film's normal, $x$. The electric current $\bj$ is applied in the $y$ direction and an effective nonequilibrium spin accumulation $\boldsymbol{\mu}$ (according to the Edelstein effect,\cite{edelsteinJPCM95} in the presence of the $C_{\infty v}$ symmetry relative to the vertical $x$ axis) is induced along the $z$ axis: $\boldsymbol{\mu}\propto\mathbf{x}\times\mathbf{j}$. This spin accumulation can be generated by a nonmagnetic heavy-metal capping layer, substrate, or the magnetic film itself. In the case of an axial symmetry about the $z$ axis, the  spin accumulation needs to exceed the gap $\omega$ in the lowest magnon band, in order to induce a magnetic instability. (The sign of $\boldsymbol{\mu}$, furthermore, needs to be consistent with the chirality of the excited mode.)}
\label{sch}
\end{figure}

Lagrangian \eqref{L} constitutes a minimal model to describe any (strongly) collinearly-ordered magnet. Pure antiferromagnetic dynamics (i.e., the standard nonlinear $\sigma$ model\cite{auerbachBOOK94}) is recovered by setting $s\to0$, while the Landau-Lifshitz equation\cite{landauBOOKv9} for the ferromagnetic case would be obtained by $\chi\to0$. Generic dissipation can be introduced into the model through the following Rayleigh dissipation function:\cite{tserkovPRB17}
\eq{
R=\frac{\alpha}{2}(\partial_t\mathbf{n})^2+\frac{g}{2}(\partial_t\mathbf{n}-\boldsymbol{\mu}\times\mathbf{n})^2\,,
\label{R}}
which complements the above Lagrangian. $\alpha$ is the Gilbert damping constant,\cite{gilbertIEEEM04} which describes the viscosity of the reorientational order-parameter dynamics, and $g$ is the spin-mixing conductance,\cite{tserkovRMP05} which describes the dissipative coupling between the spin accumulation $\boldsymbol{\mu}$ (here induced by the Edelstein/spin Hall effect) and magnetic dynamics. For a Rashba-type, i.e., $C_{\infty v}$, symmetry breaking normal to a film's ($yz$) plane, for example: $\boldsymbol{\mu}\propto\mathbf{x}\times\bj$, in terms of the applied current density $\bj$. See Fig.~\ref{sch} for a schematic.

Rayleigh function \rf{R} boils down to the dissipative torque (i.e., the net angular momentum transfer onto the collective magnetic dynamics from the nonmagnetic/incoherent degrees of freedom)
\eq{
\boldsymbol{\tau}\equiv-\mathbf{n}\times\frac{\partial R}{\partial{\partial_t\mathbf{n}}}=\mathbf{n}\times(g\boldsymbol{\mu}\times\mathbf{n}-\alpha\partial_t\mathbf{n})\,,
\label{tau}}
where $\alpha\to\alpha+g$ is henceforth the total effective damping, including also the spin-mixing-conductance contribution (known as spin pumping \cite{tserkovRMP05}). According to the Fig.~\ref{sch} schematic, we will set $\boldsymbol{\mu}\to\mathbf{z}$, lumping the current and the effective spin Hall angle,\cite{tserkovPRB14} which determine $\mu$, into $g$ (whose sign thus depends on the current direction). Other torques can be added to Eq.~\eqref{tau}, if the structural symmetries are reduced further\cite{zarzuelaPRB17} (depending on the details of the crystal and device structure), as in Ref.~\onlinecite{zeleznyPRL14}, for example. We are, furthermore, omitting some of the other torques allowed in the present high-symmetry case\cite{garelloNATN13} (like the field-like torque $\propto\bn\times\boldsymbol{\mu}$, and the less common torques $\propto n_z\bn\times\bj$, and $\propto n_z\bn\times\bj\times\bn$), which are typically less important for large-angle reorientational dynamics.

Minimizing the Lagrangian \eqref{L} with respect to $\mathbf{m}$, we find the usual expression,\cite{auerbachBOOK94}
\eq{
\mathbf{m}=\chi\mathbf{n}\times(\partial_t\mathbf{n}-\mathbf{n}\times\mathbf{b})\,,
\label{m}}
for the transverse spin density. The other Euler-Lagrange-Rayleigh equation gives the equation of motion for the total spin density:
\eq{
(\partial_t+\mathbf{b}\times)(s\mathbf{n}+\mathbf{m})+\mathbf{n}\times\partial_\mathbf{n}\mathcal{E}=\boldsymbol{\tau}\,.
\label{n}}
Setting $\chi\to0$ recovers the Landau-Lifshitz-Gilbert equation with the (damping-like) spin Hall torque, while setting instead $s\to0$ gives the standard $\sigma$-model equation for the N{\'e}el order (including spin-transfer torque and relaxation). The expression for the torque, Eq.~\eqref{tau}, is the same in both cases.\cite{tserkovPRB17} The equation of motion \eqref{n} describes Larmor-type dynamics of the total spin density, $s\mathbf{n}+\mathbf{m}$, in the presence of various torques: Zeeman, anisotropy, spin-transfer, damping, etc.\\

\textit{Easy-axis anisotropy.}|The above equations of motion, Eqs.~\eqref{m} and \eqref{n}, establish the general starting point for a collinear-order spin dynamics. As a simple illustrative example, let us now specialize it to analyze stability of the fixed points along the $\pm z$ orientations, if $z$ is the easy axis, i.e.,
\eq{
\mathcal{E}(\mathbf{n})=-\frac{Kn_z^2}{2}\,,
}
with $K>0$. The two coupled equations of motion (setting $\mathbf{b}\to0$, hereafter) are
\eq{\al{
\partial_t\mathbf{n}&=\frac{1}{\chi}\mathbf{m}\times\mathbf{n}\,,\\
\partial_t\mathbf{m}&=\mathbf{n}\times\left[Kn_z\mathbf{z}+g\mathbf{z}\times\mathbf{n}+\frac{1}{\chi}(s\mathbf{m}+\alpha\mathbf{n}\times\mathbf{m})\right]\,,
\label{nm}}}
which could be integrated up, starting from an arbitrary initial configuration, with $\mathbf{m}\perp\mathbf{n}$. Linearizing these equations close to the two equilibria, $\mathbf{n}\to\pm\mathbf{z}$, and switching to the complex notation: $n\equiv n_x+in_y$ and $m\equiv m_x+im_y$, we get
\eq{
\partial_tn=\mp\frac{i}{\chi}m~~~{\rm and}~~~\partial_t m=\mp(g+iK)n-\frac{\alpha\mp is}{\chi}m\,.
\label{nml}}
Writing these as $\partial_t\vec{s}=-\hat{A}\vec{s}$, where $\vec{s}\equiv(n,m)$ and $\hat{A}$ is the associated response matrix (that we can think of as a non-Hermitian Hamiltonian describing the small-angle dynamics), which is read out from the linearized equations \eqref{nml}, we are interested in the eigenvalue of $\hat{A}$ with the smallest real part $\lambda$. $\lambda<0$ would signal a (spin-torque-induced) instability.

This smaller (real part of the) eigenvalue is given by
\eq{
\lambda=\frac{\alpha}{2\chi}-{\rm Re}\sqrt{\left(\frac{\alpha\mp is}{2\chi}\right)^2-\frac{K-ig}{\chi}}\,,
\label{l}}
where the square root is evaluated for its principal value (yielding the nonnegative real part). As a sanity check, we verify that $\lambda\geq0$ if $g=0$.\\

\textit{Ferromagnetic case.}|For the ferromagnetic (F) case, $\chi\to0$, we get
\eq{
\lambda_{\rm F}={\rm Re}\frac{K-ig}{\alpha\mp is}=\frac{\alpha K\pm gs}{\alpha^2+s^2}\,.
}
The instability sets in when $\lambda_{\rm F}\to0$, which corresponds to
\eq{
g=\mp\alpha\frac{K}{s}=\mp\alpha\omega_{\rm F}\,,
}
where $\omega_{\rm F}\equiv K/s$ is the ferromagnetic resonance frequency. For the equilibrium $\mathbf{n}$ oriented along the $\pm z$ axis, we thus need a negative (positive) torque $g$ in order to trigger the F instability. This results in the familiar unidirectional switching of the F orientation towards the $\mp$ orientation (which is then stable against the torque). The above threshold makes sense thermodynamically: In the absence of intrinsic damping (so that $\alpha$ is determined by the spin-mixing conductance $g$), the spin accumulation $\boldsymbol{\mu}$ must overcome the intrinsic gap in the magnon spectrum, i.e., $\omega_{\rm F}$, while oriented parallel to the spin angular momentum of individual magnons.\cite{benderPRL12,*benderPRB14} A finite intrinsic damping raises the instability threshold further.\\

\textit{Antiferromagnetic case.}|For the antiferromagnetic (AF) case, $s\to0$:
\eq{\al{
\lambda_{\rm AF}&=\frac{\alpha}{2\chi}-{\rm Re}\sqrt{\left(\frac{\alpha}{2\chi}\right)^2-\frac{K-ig}{\chi}}\\
&\approx\frac{\alpha}{2\chi}-{\rm Re}\sqrt{-\frac{K-ig}{\chi}}=\frac{\alpha}{2\chi}+{\rm Im}\sqrt{\frac{K-i|g|}{\chi}}\\
&\approx\frac{\alpha}{2\chi}-\frac{|g|}{2\sqrt{K\chi}}\,,
}}
where the approximations are based on the assumptions that $\alpha\ll\sqrt{\chi K}$ (which physically corresponds to the quality factor of the AF resonance $Q\gg1$) and $|g|\ll K$ (which in fact follows from $\alpha\ll\sqrt{\chi K}$, at the onset of the instability: $\lambda_{\rm AF}\to0$). Independently of the sign of the torque $g$, we reach the instability at\cite{chenPRL18}
\eq{
|g|=\alpha\sqrt{\frac{K}{\chi}}=\alpha\,\omega_{\rm AF}\,.
}
$\omega_{\rm AF}\equiv\sqrt{K/\chi}$ is the AF resonance frequency. As in the F case, this result makes sense thermodynamically: In the absence of intrinsic damping (so that $\alpha$ is determined by the spin pumping $\propto g$), the spin accumulation $\boldsymbol{\mu}$ must overcome the intrinsic gap in the magnon spectrum, i.e., $\omega_{\rm AF}$. The sign of $g$ determines which of the two magnon branches of the spectrum goes unstable. Beyond the threshold of instability, the order parameter $\mathbf{n}$ reaches a stable precession within the $xy$ plane, which we obtain according to Eqs.~\eqref{nm} (with $s\to0$) as
\eq{
\partial_t\mathbf{n}=\frac{g}{\alpha}\mathbf{z}\times\mathbf{n}\,,~~~\mathbf{m}=\frac{\chi g}{\alpha}\mathbf{z}\,.
\label{nmc}}
For this trajectory, the torque \eqref{tau} vanishes, so that the work done on the magnetic dynamics by the spin transfer $g$ is dissipated by Gilbert damping $\alpha$. The corresponding precession frequency is $\omega=g/\alpha$ (recall that $g$ is proportional to the applied electric current).\\

\textit{Spin-glass case.}|For a spin glass, we expect the dynamics similar to that of the AF case above.\cite{ochoaPRB18} Namely, beyond the threshold current set by anisotropies (which may be randomized and thus reduced by disorder), the SO(3)-valued state variable (which is rooted in the Edwards-Anderson order parameter\cite{edwardsJPF75}), will precess around the $z$ axis at the frequency
\eq{
\omega_{\rm SG}=\frac{g}{\alpha}\,,
\label{sg}}
where $g$ is the appropriate spin-mixing conductance, which is closely analogous to the collinear case,\cite{tserkovPRB17} and $\alpha$ is the spin-glass Gilbert damping. At the same time, a nonequilibrium spin polarization $\mathbf{m}$ builds up, as in Eq.~\eqref{nmc}, determined by the spin-glass susceptibility $\chi$. This spin polarization, with magnitude $\chi\omega_{\rm SG}$, makes sense from the rotating-frame perspective.\\

\textit{Ferrimagnetic case.}|Since the F and the AF cases showed qualitatively different behavior at the instability threshold|namely, the F order tends to undergo a unipolar switching along the easy axis, while the AF order settles at an equatorial limit cycle perpendicular to the easy axis|it is interesting to study the interplay between these different tendencies in \textit{ferri}magnets. This, in particular, can provide us a blueprint for driven dynamics in complex systems with competing magnetic orders.

We thus return to discuss the instability threshold in the general (ferrimagnetic) case, Eq.~\eqref{l}, with both $\chi$ and $s$ being nonzero. In analogy to the above F and AF cases, we expect the threshold to be reached when $g$ approaches $\alpha\omega$, in terms of the respective resonance frequency $\omega$ (supposing the overall damping is weak, so that the quality factor of the dynamics is $\gg1$), which is easily verified also directly from Eq.~\eqref{l}. The normal-mode frequencies for solutions $\propto e^{i\omega t}$ are obtained from Eqs.~\eqref{nml}, after setting $\alpha,g\to0$:
\eq{
\omega=\sigma\sqrt{\frac{K}{\chi}+\frac{s^2}{4\chi^2}}\pm\frac{s}{2\chi}\,,
\label{w}}
with two magnon branches labelled by $\sigma\to\pm$. As before, the other $\pm$ in Eq.~\rf{w} stands for the initial $\pm z$ orientation of the order parameter.
The positive (negative) frequency (corresponding here to $\sigma\to\pm$) is associated with magnons carrying spin angular momentum $\pm\hbar$, thus requiring a positive (negative) torque $g$ to reach the threshold. Using Eq.~\eqref{w}, we recover the F limit when $s/\sqrt{\chi K}\to\infty$ and the AF limit when $s/\sqrt{\chi K}\to0$.

\begin{figure}[!t]
\begin{center}\includegraphics[width=\linewidth]{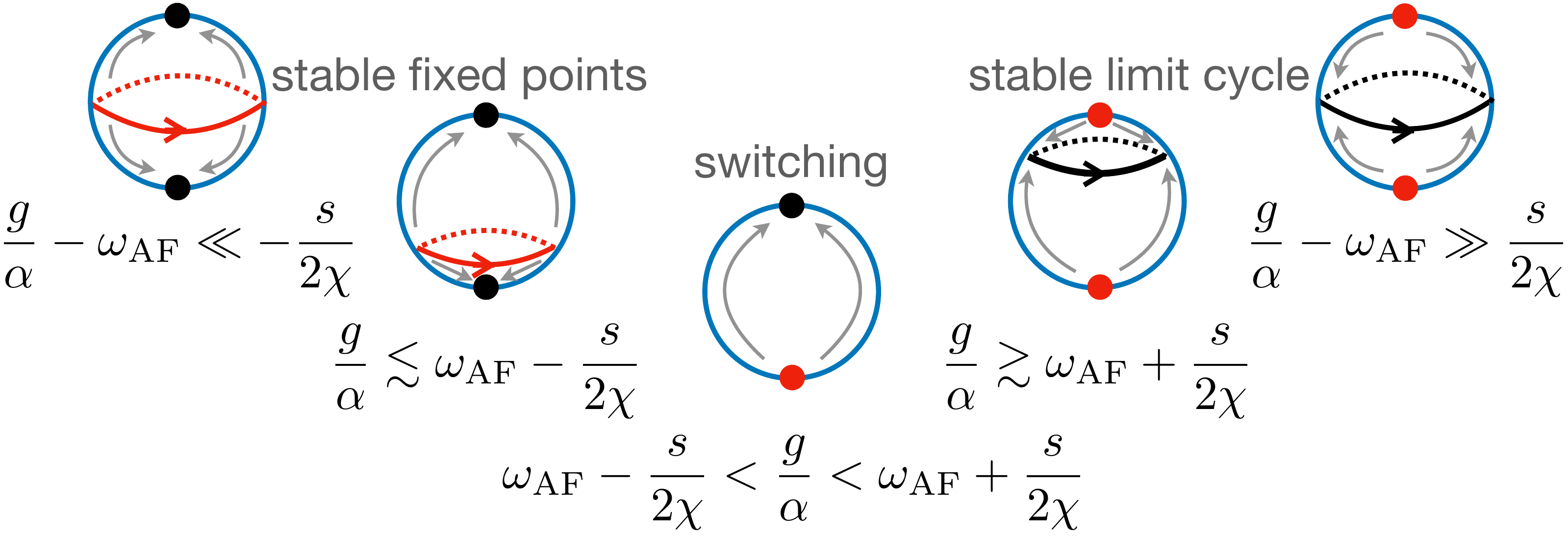}\end{center}
\caption{Evolution of the dynamical state for the order parmeter $\bn(t)$ according to Eq.~\eqref{nm}, in the ferrimagnetic case [cf. Eq.~\eqref{ws}]. Stable (unstable) fixed points and limit cycles are shown in black (red). For small positive $g$, two stable fixed points at $n_z\to\pm1$ coexist with an unstable limit cycle (in the lower hemisphere) near the $xy$ plane. For large $g$, these fixed points become unstable, while the limit cycle stabilizes (in the upper hemisphere), analogously to the antiferromagnetic case. At intermediate values of $g$, only one fixed point is stable, as in the ferromagnetic case.}
\label{dyn}
\end{figure}

Let us now start from the AF limit, and consider the weakly ferrimagnetic case, $s\ll\sqrt{\chi K}$. The normal-mode frequencies \eqref{w} (corresponding in the continuum description to the gaps of the respective magnon branches) then become
\eq{
\omega\approx\sigma\sqrt{\frac{K}{\chi}}\pm\frac{s}{2\chi}=\sigma\omega_{\rm AF}\pm\frac{s}{2\chi}\,.
\label{ws}}
When $s$ is increased, the lower of these two frequencies would eventually approach the ferromagnetic resonance frequency ($\omega_{\rm F}=K/s$), according to Eq.~\eqref{w}. Equation~\eqref{ws} shows that the threshold of instability depends on the orientation of $\mathbf{n}$ and the sign of $g$. For instance, when $g>0$, the $\omega>0$ branch is excited and the critical current corresponds to the $\sigma\to+$ (i.e., spin-up magnon) mode, which is now split by $s/\chi$, depending on the initial up or down orientation of $\mathbf{n}$ along the $z$ axis [denoted by $\pm$ in Eq.~\eqref{ws}]. We thus conclude that when $g/\alpha<\omega_{\rm AF}-s/2\chi$, both $\mathbf{n}\to\pm\mathbf{z}$ are stable fixed points; when $\omega_{\rm AF}-s/2\chi<g/\alpha<\omega_{\rm AF}+s/2\chi$, the $\mathbf{n}\to\mathbf{z}$ is stable, while $\mathbf{n}\to-\mathbf{z}$ is unstable (allowing for the unidirectional switching, as in the ferromagnet); and when $g/\alpha>\omega_{\rm AF}+s/2\chi$, both are unstable (resulting in a stable precessional state, as in the AF case, albeit somewhat canted out of the $xy$ plane). In Fig.~\ref{dyn}, we show a schematic of how this dynamical system evolves as a function of $g>0$.\\

\textit{Summary.}|This Letter aims to establish basic aspects and intuition for thinking about spin-torque instabilities and the ensuing dynamics of the ordered magnets and spin glasses, under the generic dissipative torque \rf{tau}. To summarize: Beyond an instability threshold, which is determined by the anisotropies, the order parameter generally has a tendency to precess according to the right-hand rule around the spin accumulation $\boldsymbol{\mu}\propto\mathbf{z}$. This is seen from the torque \eqref{tau}, which imparts positive work $\propto\mu$ for the right-hand precession. As a result of the instability, the order parameter either switches (cf. the ferromagnetic case) or precesses steadily (cf. the antiferromagnetic and spin-glass phases), in which case the steady input of work is dissipated by Gilbert damping. The ferrimagnetic case (with both $s,\chi\neq0$) combines both of these aspects, as sketched in Fig.~\ref{dyn}.

Generally, when under the action of the damping torque \rf{tau}, we expect for the F spin order to tend to point along the applied spin accumulation (depending on its sign). In this case, the magnons associated with a potential destabilization of the order, which carry spin opposite to the order parameter, are thermodynamically biased to be ejected by the spin accumulation.\cite{benderPRL12} The AF order, on the other hand, tends to dynamically orient normal to the spin accumulation (independent of its sign). Such order-parameter reorientations can result in characteristic electrical magnetoresistive or X-ray dichroism signatures.\cite{wadleySCI16,*wadleyNATN18,chenPRL18,baldratiPRL19} In the multidomain case, the possible reorientation of domains depends crucially on the dynamics within the domain walls, whose motion is likewise driven by a positive gain of the work by the spin torque $\propto\mu$ (thus being sensitive to the chiral structure of the moving domain walls\cite{baldratiPRL19}).

It is interesting to consider a situation, in which a torque-driven precessional state is reached in a spin glass (SG) coexisting with another magnetic order.\cite{manivCM20} In the case of a (coarse-grained) rotational symmetry about the $z$ axis, furthermore, the threshold for SG dynamics could be low. The steady-state SG frequency is given by Eq.~\eqref{sg}. In the rotating frame of reference, other degrees of freedom would get polarized by the fictitious Zeeman field of $\omega_{\rm SG}$ along the rotation (i.e., $z$) axis. This, in turn, would facilitate the switching towards the $z$ axis in the F case or precession towards the $xy$ plane in the AF case. The details would depend on the exact interaction of the coexisting SG and collinear magnetic components (and could in principle be handled with our Lagrangian-Rayleigh approach,\cite{tserkovPRB17} subject to a decomposition of the magnetic state into the SG and ferrimagnetic variables). In this sense, the SG dynamics can be viewed as enhancing the torque transfer from the electronic to the magnetic degrees of freedom.\cite{manivCM20}

Finally, we wish to emphasize that, in our analysis, we have retained only the most generic Slonczewski-like\cite{slonczewskiPRB89} torque \rf{tau} due to the Edelstein/spin Hall effect, as a means to pump energy into magnetic precession and cause dynamic instabilities. Other types of torques, as discussed below Eq.~\rf{tau}, may need to be included in more specialized cases, especially if the symmetries are reduced by crystalline order.\\

The work was supported by the U.S. Department of Energy, Office of Basic Energy Sciences under Award No. DE-SC0012190.\\

Data sharing is not applicable to this article as no new data were created or analyzed in this study.

\end{document}